\documentclass[aps,prl,twocolumn,showpacs,superscriptaddress,floatfix]{revtex4}

\usepackage{graphicx} % Include figure files       
\usepackage{dcolumn} % Align table columns on decimal point       
\usepackage{bm} % bold math       
\def\FigFactor{0.5}
\def\sqrtsNN{\mbox{$\sqrt{s_\mathrm{NN}}$}}

\def\GeVc{\mbox{$\mathrm{GeV}/c$}}

\def\lt{\mbox{$<$}}

\newcommand{ \be }{\begin{equation}}    
\newcommand{ \ee }{\end{equation}}    
\newcommand{ \bea }{\begin{eqnarray}}    
\newcommand{ \eea }{\end{eqnarray}}    
\newcommand{ \la }{\langle}    
\newcommand{ \ra }{\rangle}    
    
\newcommand{ \mean }[1]{\left\langle #1 \right\rangle} 
\usepackage{color}

\begin{document}       
%\voffset=0.5 in   % needed for 8 1/2 X 11 paper if dvips
% does not have -t letter 
       
\title{    
%\begin{flushright}    
%{{\small \sl \version \\ \today \\ for Phys. Rev. Letters }}    
%\end{flushright}    
Azimuthal anisotropy at RHIC: the first and fourth harmonics }

%============================================================================

%\author{The STAR Collaboration}
%\documentclass[aps,prl,superscriptaddress]{revtex4}
%\begin{document}
%\title{STAR Collaboration Author List - July 7, 2003}
%STAR author list in APS format for use with revtex4 i.e. for PRL, PRC etc.
%Force affiliation order to be alphabetic, otherwise they come out in the author order

\affiliation{Argonne National Laboratory, Argonne, Illinois 60439}
\affiliation{Brookhaven National Laboratory, Upton, New York 11973}
\affiliation{University of Birmingham, Birmingham, United Kingdom}
\affiliation{University of California, Berkeley, California 94720}
\affiliation{University of California, Davis, California 95616}
\affiliation{University of California, Los Angeles, California 90095}
\affiliation{Carnegie Mellon University, Pittsburgh, Pennsylvania 15213}
\affiliation{Creighton University, Omaha, Nebraska 68178}
\affiliation{Nuclear Physics Institute AS CR, \v{R}e\v{z}/Prague, Czech Republic}
\affiliation{Laboratory for High Energy (JINR), Dubna, Russia}
\affiliation{Particle Physics Laboratory (JINR), Dubna, Russia}
\affiliation{University of Frankfurt, Frankfurt, Germany}
\affiliation{Indiana University, Bloomington, Indiana 47408}
\affiliation{Insitute  of Physics, Bhubaneswar 751005, India}
\affiliation{Institut de Recherches Subatomiques, Strasbourg, France}
\affiliation{University of Jammu, Jammu 180001, India}
\affiliation{Kent State University, Kent, Ohio 44242}
\affiliation{Lawrence Berkeley National Laboratory, Berkeley, California 94720}\affiliation{Max-Planck-Institut f\"ur Physik, Munich, Germany}
\affiliation{Michigan State University, East Lansing, Michigan 48824}
\affiliation{Moscow Engineering Physics Institute, Moscow Russia}
\affiliation{City College of New York, New York City, New York 10031}
\affiliation{NIKHEF, Amsterdam, The Netherlands}
\affiliation{Ohio State University, Columbus, Ohio 43210}
\affiliation{Panjab University, Chandigarh 160014, India}
\affiliation{Pennsylvania State University, University Park, Pennsylvania 16802}
\affiliation{Institute of High Energy Physics, Protvino, Russia}
\affiliation{Purdue University, West Lafayette, Indiana 47907}
\affiliation{University of Rajasthan, Jaipur 302004, India}
\affiliation{Rice University, Houston, Texas 77251}
\affiliation{Universidade de Sao Paulo, Sao Paulo, Brazil}
\affiliation{University of Science \& Technology of China, Anhui 230027, China}
\affiliation{Shanghai Institute of Nuclear Research, Shanghai 201800, P.R. China}
\affiliation{SUBATECH, Nantes, France}
\affiliation{Texas A\&M, College Station, Texas 77843}
\affiliation{University of Texas, Austin, Texas 78712}
\affiliation{Valparaiso University, Valparaiso, Indiana 46383}
\affiliation{Variable Energy Cyclotron Centre, Kolkata 700064, India}
\affiliation{Warsaw University of Technology, Warsaw, Poland}
\affiliation{University of Washington, Seattle, Washington 98195}
\affiliation{Wayne State University, Detroit, Michigan 48201}
\affiliation{Institute of Particle Physics, CCNU (HZNU), Wuhan, 430079 China}
\affiliation{Yale University, New Haven, Connecticut 06520}
\affiliation{University of Zagreb, Zagreb, HR-10002, Croatia}
%Authors 1 per line follow by affiliation, three authors have two affiliations
\author{J.~Adams}\affiliation{University of Birmingham, Birmingham, United Kingdom}
\author{C.~Adler}\affiliation{University of Frankfurt, Frankfurt, Germany}
\author{M.M.~Aggarwal}\affiliation{Panjab University, Chandigarh 160014, India}
\author{Z.~Ahammed}\affiliation{Variable Energy Cyclotron Centre, Kolkata 700064, India}
\author{J.~Amonett}\affiliation{Kent State University, Kent, Ohio 44242}
\author{B.D.~Anderson}\affiliation{Kent State University, Kent, Ohio 44242}
\author{M.~Anderson}\affiliation{University of California, Davis, California 95616}
\author{D.~Arkhipkin}\affiliation{Particle Physics Laboratory (JINR), Dubna, Russia}
\author{G.S.~Averichev}\affiliation{Laboratory for High Energy (JINR), Dubna, Russia}
\author{S.K.~Badyal}\affiliation{University of Jammu, Jammu 180001, India}
% ...

\collaboration{STAR Collaboration}\homepage{www.star.bnl.gov}\noaffiliation

%\maketitle
%\end{document}

%=============================================================================

\date{\today}

\begin{abstract}
We report the first observations of the first harmonic (directed flow,
$v_1$), and the fourth harmonic ($v_4$), in the azimuthal distribution
of particles with respect to the reaction plane in Au+Au collisions at
the Relativistic Heavy Ion Collider (RHIC).  Both measurements were
done taking advantage of the large elliptic flow ($v_2$) generated at
RHIC. From the correlation of $v_2$ with $v_1$ it is determined that
$v_2$ is positive, or {\it in-plane}. The integrated $v_4$ is about a
factor of 10 smaller than $v_2$. For the sixth ($v_6$) and eighth
($v_8$) harmonics upper limits on the magnitudes are reported.
\end{abstract}


\maketitle

%==============================================================================

Anisotropic flow, an anisotropy of the particle azimuthal distribution
in momentum space with respect to the reaction plane, is a sensitive
tool in the quest for the quark-gluon plasma and the understanding of
bulk properties of the system created in ultrarelativistic nuclear
collisions~\cite{QM}. It is commonly studied by measuring the Fourier
harmonics ($v_n$) of this distribution~\cite{Methods}.  Elliptic flow,
$v_2$, is well studied at RHIC~\cite{STARcumulants,v2,PHOBOS} and is
thought to reflect conditions from the early time of the collision.
Directed flow, $v_1$, was discovered almost 20 years
ago~\cite{CollectiveFlow} and has been extensively studied and
reviewed at lower beam energies~\cite{reviews}.  At RHIC energies
directed flow in the central rapidity region reflects important
features of the system evolution from its initial conditions.  $v_1$
is predicted to be small near midrapidity with almost no dependence on
pseudorapidity. However, it could exhibit a characteristic
"wiggle"~\cite{wiggle}, depending on the baryon stopping and
production mechanisms as well as strong space-momentum correlations in
the system's evolution.  A similar rapidity dependence of directed
flow could develop due to a change in the matter compressibility if a
quark-gluon plasma is formed~\cite{antiflow,third-component}. It
results in the so-called third flow component~\cite{third-component}
or ``anti flow"~\cite{antiflow} component in the expansion of the
matter. This expansion direction is opposite to the normal directed
flow.  $v_1$ has not previously been reported at RHIC.

The importance of the higher harmonics in understanding the initial
configuration and the system evolution has been emphasized~\cite{v4}.
Recently, Kolb~\cite{Kolb_v4} reported that the magnitude and even the
sign of $v_4$ are more sensitive than $v_2$ to initial conditions in
the hydrodynamic calculations.  Those higher harmonics reflect the
details of the initial configuration geometry.  Besides one early
measurement at the AGS~\cite{E877PRL94}, reports of higher harmonics
have not previously been published.

{\it Experiment---} The data come from the reaction Au~+~Au at
$\sqrtsNN = 200$ GeV. The STAR detector~\cite{STAR} main time
projection chamber (TPC~\cite{STARTPC}) and two forward TPCs
(FTPC~\cite{STARFTPC}) were used in the analysis. For the higher
harmonics 2 million events in the main TPC were analyzed.  For the
first harmonic analysis there were 70 thousand events available which
included the FTPCs.

In this analysis the main TPC covered pseudorapidity ($\eta$) from
--1.2 to 1.2, while two FTPCs covered --4.2 to --2.4 and 2.4 to 4.2.
The low transverse momentum ($p_t$) cutoff was 0.15 \GeVc. In the
present work all charged particles were analyzed, regardless of their
particle type. The centrality definition in this paper is the same as
used previously by STAR~\cite{centrality}. The errors presented in the
figures are statistical.

{\it Analysis---} The difficulties in studying directed flow are
that the signal is small and the non-flow contribution to the
two-particle azimuthal correlations can be comparable or even larger
than the correlations due to flow.  To suppress the non-flow effects
the current analysis uses the knowledge about the reaction plane
derived from the large elliptic flow.  One method for eliminating the
non-flow contribution in a case when the reaction plane is known was
proposed in~\cite{Methods}.  It was noted that while the correlations
of the components of the (first harmonic) flow vectors in the reaction
plane contain both flow and non-flow contributions, the correlations
of the components perpendicular to the reaction plane contain only
non-flow contributions. Then the difference yields the flow
contribution.  Correlating the azimuthal angles of two particles
($\phi_a,
\phi_b$), and using the event plane determined by elliptic flow
($\Psi_2$) one gets:
\bea
&& \la \cos(\phi_a - \Psi_{2}) \cos(\phi_b - \Psi_{2}) \\
&& \;\;\;\;\;\;\;\; \;\;\;\;\;\;\;\; \;\;\;\;\;\;\;\;
 - \sin(\phi_a - \Psi_{2}) \sin(\phi_b - \Psi_{2}) \ra \nonumber \\
&&= \la  \cos(\phi_a +\phi_b -2 \Psi_{2}) \ra 
\approx v_{1,a}v_{1,b} \la \cos(2(\Psi_2-\Psi_{RP})) \ra, \nonumber
\eea
where $\Psi_{RP}$ is the azimuthal angle of the reaction plane.  If
only one particle is used to determine the second harmonic event plane
this expression reduces to
\be
\la  \cos(\phi_a +\phi_b -2 \phi_{c}) \ra
\approx v_{1,a} v_{1,b} v_{2,c},
\label{cos3part}
\ee
which is the basic formula of the three-particle correlation method of
Borghini, Dinh, and Ollitrault~\cite{v1{3}}.  The analysis of directed
flow in this paper is performed using this three-particle cumulant
method~\cite{v1{3}}.  The analyses for $v_4$, $v_6$, and $v_8$ were
done relative to the second harmonic event plane using the method
described in Refs.~\cite{Methods,Olli97}, with the event plane
resolution calculated from Ref.~\cite{Methods} equation~11 with $k=2,
3,$ or $4$. Note that this approach in many aspects is very similar to
the analysis of directed flow described above as it also involves
three (for $v_4$, and four for $v_6$) particle correlations. For
example, for the fourth harmonic flow, (approximately, for the exact
relations actually used in the analysis, see~\cite{Methods})
\be
\la  \cos(4\phi  -4 \Psi_{2}) \ra \approx  v_{2}^2 v_{4} \, N/2,
\label{cos4}
\ee
where $N$ is the total number of particles used to determine the
second harmonic event plane.  This expression should be compared to
Eq.~(\ref{cos3part}).  Results obtained with this method we designate
by $v_4\{EP_2\}$.  The analysis for $v_4$ was also done with
three-particle cumulants~\cite{cumulants} by measuring $\la
\cos(2\phi_a+2\phi_b-4\phi_c) \ra$.

%-----------------------------------------------------------------------
\begin{figure}[ht]
\resizebox{\FigFactor\textwidth}{!}{\includegraphics{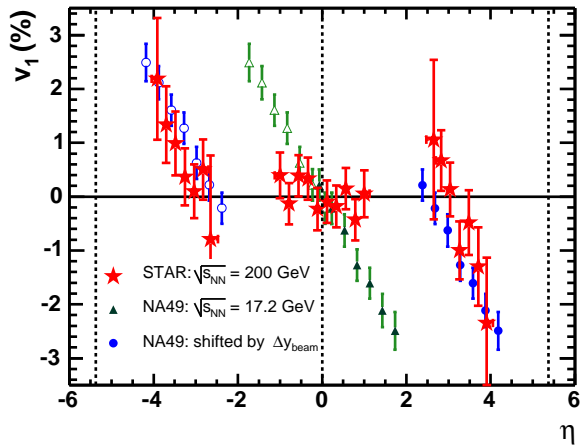}}
\caption{
(color online). The values of $v_1$ (stars) for charged particles for
10\% to 70\% centrality plotted as a function of pseudorapidity. Also
shown are the results from NA49 (triangles) for pions from 158$A$ GeV
Pb + Pb midcentral (12.5\% to 33.5\%) collisions plotted as a function
of rapidity. The open points have been reflected about
midrapidity. The NA49 points have also been shifted (circles) plus or
minus by the difference in the beam rapidities of the two
accelerators. The dashed lines indicate midrapidity and RHIC beam
rapidity. Both results are from analyses involving three-particle
cumulants, $v_{1}\{3\}$.
\label{v1}}
\end{figure}
%-----------------------------------------------------------------------

{\it $v_1$ results---} Fig.~\ref{v1} shows the results in comparison
to the lower beam energy data of NA49~\cite{NA49}. The NA49 data are
also replotted so as to be at the same distance from beam
rapidity~\footnote{For the STAR data the beam rapidity was taken as
5.37 and for NA49 as 2.92 in the center-of-mass frame.} as the STAR
results. The RHIC $v_1(\eta)$ results differ greatly from the
unshifted SPS data in that they are flat near midrapidity and only
become significant at the highest rapidities measured. However, when
plotted in the projectile frame relative to their respective beam
rapidities, they look similar.
%~\footnote{Assuming the STAR charged
%particles are pions and plotting versus rapidity instead of
%pseudorapidity makes no perceptible difference in the graph. However,
%it should be noted that the NA49 proton curves, which would contribute
%to the charged particles, are flatter near midrapidity.}.
It should be noted that at the SPS energies of 40$A$~GeV and
158$A$~GeV~\cite{NA49}, this $y - y_{beam}$ scaling does not work, but
$y/y_{beam}$ scaling does.  In the pseudorapidity region $|\eta|<1.2$,
$v_1(\eta)$ is approximately flat with a slope of $(-0.25 \pm 0.27
(stat) )$\% per unit of pseudorapidity, which is consistent with
predictions~\cite{wiggle, third-component, antiflow}.

Note that the sign of $v_1$ is undetermined because $v_1$ enters as
the square in Eq.~(\ref{cos3part}).  We have plotted $v_1$ in the
positive hemisphere going negative toward beam rapidity as it does at
the lower beam energy. In the NA49 analysis~\cite{NA49} the sign of
$v_1$ had been determined by defining $v_1$ for protons near beam
rapidity to be positive for peripheral collisions.  On the other hand,
since the measured correlation of Eq.~(\ref{cos3part}) is
positive, we can conclude that we have measured the sign of $v_2$ to
be positive. While the absolute values of $v_2$ at RHIC are well
determined~\cite{STARcumulants,v2,PHOBOS} this is the first direct
indication that the elliptic flow at RHIC is {\it in-plane}.

%-----------------------------------------------------------------------
\begin{figure}[ht]
\resizebox{\FigFactor\textwidth}{!}{\includegraphics{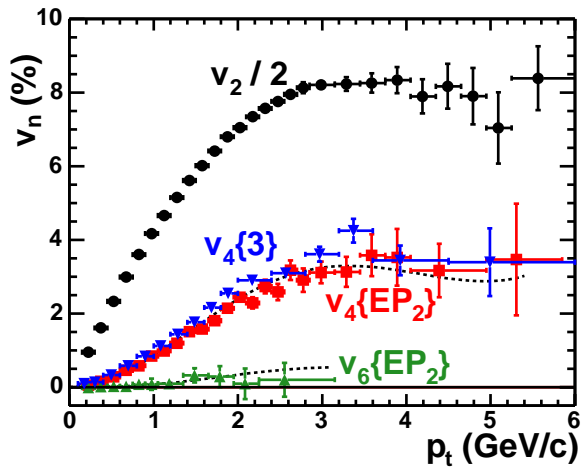}}
\caption{
(color online). The minimum bias values of $v_2$, $v_4$, and $v_6$
with respect to the second harmonic event plane as a function of $p_t$
for $| \eta | \lt 1.2$. The $v_2$ values have been divided by a factor
of two to fit on scale. Also shown are the three particle cumulant
values (triangles) for $v_4$ ($v_4\{3\}$). The dashed curves are $1.2
\cdot v_2^2$ and $1.2 \cdot v_2^3$.
\label{v(pt)}}
\end{figure}
%-----------------------------------------------------------------------

{\it $v_4$ results---} The results as a function of $p_t$ are shown in
Fig.~\ref{v(pt)} for minimum bias collisions ($0 - 80$\%
centrality). Shown for $v_4$ are both the analysis relative to the
second harmonic event plane, $v_4\{EP_2\}$, and the three-particle
cumulant, $v_4\{3\}$. Both methods determine the sign of $v_4$ to be
positive. As a function of $p_t$, $v_4$ rises more slowly from the
origin than $v_2$, but does flatten out at high $p_t$ like
$v_2$. The $v_6(p_t)$ values are consistent with zero. The
hydrodynamic calculations of Kolb~\cite{Kolb_v4} for pions from
$b=7$~fm collisions agree very well with our measured $v_4$ for
charged particles for centrality 20 to 30\%. However, he calculates
$v_6$ to be $-1.2\%$ at 2 \GeVc, while we observe in Fig.~\ref{v(pt)}
for minimum bias data that it is essentially zero. It also appears to
be zero in our data for all the individual centralities. Ollitrault
has proposed~\cite{OlliScaling} for the higher harmonics that $v_n$
might be proportional to $v_2^{n/2}$ if the $\phi$ distribution is a
smooth, slowly varying function of $\cos(2 \phi)$. In order to test
the applicability of this scaling we have also plotted $v_2^2$ and
$v_2^3$ in the figure as dashed lines. The proportionality constant
has been taken to be $1.2$ in order to fit the $v_4$ data.

Kolb~\cite{Kolb_v4} points out that for $v_2 > 10\%$, which occurs at
high $p_t$, and no other harmonics, the azimuthal distribution is
not elliptic, but becomes ``peanut'' shaped.  He calculates the amount
of $v_4$ (which looks like a four-leaf clover) needed to eliminate
this waist.  Our values of $v_4$ as a function of $p_t$ are about a
factor of two larger than needed to just eliminate this waist.

The results for $v_4$ as a function of pseudorapidity are
approximately flat in the acceptance of the main TPC ($| \eta |
\lt 1.2$) with an average value of ($0.44 \pm 0.02$)\%. However,
in the FTPCs ($2.7 \lt | \eta | \lt 4.0$) the average value is ($0.06
\pm 0.07$)\%, consistent with zero, with a two sigma upper limit of
0.2\%.  Consistent with the first observation by
PHOBOS~\cite{PHOBOS}, at $\eta =$ 3 for minimum bias collisions we
observe $v_2$ = $(3.06 \pm 0.10)\%$, which is a factor 1.8 smaller
than at midrapidity.  Thus $v_4$ seems to fall off faster at high
rapidity than $v_2$.  This faster fall off at high pseudorapidity is
also consistent with $v_4$ scaling like $v_2^2$.

%-----------------------------------------------------------------------
\begin{figure}[ht]
\resizebox{\FigFactor\textwidth}{!}{\includegraphics{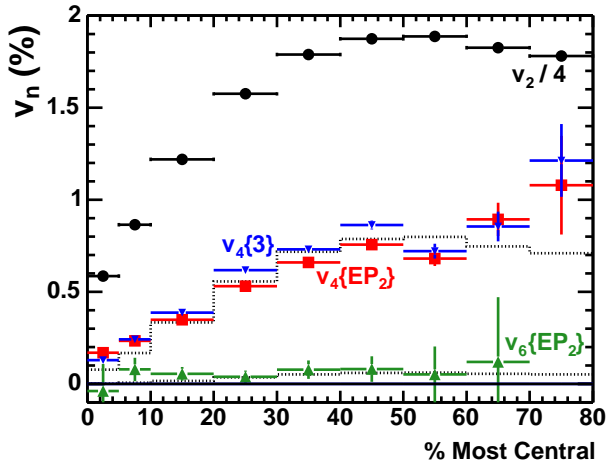}}
\caption{
(color online). The $p_t$- and $\eta$- integrated values of $v_2$,
$v_4$, and $v_6$ as a function of centrality. The $v_2$ values have
been divided by a factor of four to fit on scale. Also shown are the
three particle cumulant values for $v_4$ ($v_4\{3\}$). The dotted
histograms are $1.4 \cdot v_2^2$ and $1.4 \cdot v_2^3$.
\label{v}}
\end{figure}
%-----------------------------------------------------------------------

Fig.~\ref{v} shows the centrality dependence for $p_t$-integrated
$v_2$, $v_4$, and $v_6$ with respect to the second harmonic event
plane and also $v_4$ from three-particle cumulants ($v_4\{3\}$).  The
five-particle cumulant, $v_4\{5\}$, (not shown in the figure) is
consistent with both methods but the error bars are about two times
larger. The $v_6$ values are close to zero for all centralities.
These results are averaged over $p_t$, thus reflecting mainly the low
$p_t$ region where the yield is large, and also averaged over $\eta$
for the midrapidity region accessible to the STAR TPC ($| \eta | \lt
1.2$). To again test the applicability of $v_2^{n/2}$ scaling we have
also plotted $v_2^2$ and $v_2^3$ in the figure as dotted
histograms. The proportionality constant has been taken to be $1.4$ to
approximately fit the $v_4$ data. The larger constant here compared to
that used in Fig.~\ref{v(pt)} is understood as coming from the use of
the square of the average instead of the average of the square, and
because the integrated values yield-weight low $p_t$ more, where the
best factor is slightly larger.

The $v_n\{EP_2\}$ values averaged over $p_t$ and $\eta$ ($ |
\eta | \lt 1.2 $), and also centrality (minimum bias, $0 - 80 \%$),
are (in percent) $v_2 = 5.18 \pm 0.005$, $v_4 = 0.44 \pm 0.009$, $v_6
= 0.043 \pm 0.037$, and $v_8 = - 0.06 \pm 0.14$. Since $v_6$ is
essentially zero, we place a two sigma upper limit on $v_6$ of
0.1\%. Also, $v_8$ is zero, but the error is larger because the
sensitivity decreases as the harmonic order increases.

{\it Systematic uncertainties---} In both approaches, $v_4\{3\}$ and
$v_4\{EP_2\}$, the non-flow effects are suppressed compared to the
case where the fourth harmonic event plane is used. The remaining
non-flow correlations, along with event-by-event flow fluctuations,
are thought to be the major contributors to the systematic
uncertainties. Background from secondary particles is expected to be
less than 15\%, and remaining acceptance effects are measured to be
very small. All errors and limits quoted so far are statistical, and
should be increased by the systematic uncertainties below.

From non-flow effects we estimate the relative systematic uncertainty
in $v_4\{3\}$ to be about 20\%. The largest contribution comes from
situations in Eq.~(\ref{cos4}) where one particle is correlated with
one of the other particles due to non-flow, and with the third
particle via flow. Our estimate is based on the assumption that the
entire difference in the published values~\cite{STARcumulants} of
$v_2\{EP_2\}$ and $v_2\{4\}$ is due to non-flow effects. Comparison of
$v_4\{3\}$ to $v_4\{5\}$ leads to a similar estimate for this
systematic error.

From non-flow effects we estimate the relative systematic uncertainty
in $v_1\{3\}$ also to be about 20\%. Our estimate is based on the
assumption that our two-particle correlation value of $v_1$ using only
the first harmonic event plane in the FTPCs, $v_1\{EP_1\}$, of about
3\% is entirely due to non-flow effects.

The other effect important for the comparison of our results to
theoretical calculations is event-by-event flow fluctuations. As was
discussed~\cite{STARcumulants}, flow measurements are done by two or
many particle correlations, resulting in, not $\mean{v_n}$, but
$\mean{v_n^k}^{1/k}$. If flow fluctuates event-by-event, it could lead
to a difference between these two quantities. Fluctuations in the
initial geometry of the collision at fixed impact parameter can
account for the difference between $v_2\{EP_2\}$ and
$v_2\{4\}$~\cite{STARcumulants}, and also between $v_4\{EP_4\}$ and
$v_4\{3\}$~\cite{epsilonMC}. Although the flow fluctuation
contribution to $v_4\{3\}$ is greatly reduced, it still could lead to
an effect of about a factor of 1.2 to 1.5.

{\it Conclusions---} We have presented the first measurement of $v_1$
at RHIC energies. $v_1(\eta)$ is found to be approximately flat in the
midrapidity region, which is consistent with microscopic transport
models, as well as hydrodynamical models where the flatness is
associated with the development of the expansion in the direction
opposite to the normal directed flow. Within errors we do not observe
a wiggle in $v_1(\eta)$ at midrapidity. The pseudorapidity dependence
of $v_1$ in the projectile fragmentation region is very similar to
that observed at full SPS energy. We observe a positive correlation
between the first and second harmonics, indicating that elliptic flow
is {\it in-plane}. This is the first direct measurement at RHIC of the
orientation of elliptic flow relative to the reaction plane.

We have measured $v_4$ as a function of $p_t$, $\eta$, and
centrality. We observe that $v_4$ appears to scale approximately as
$v_2^2$, as a function of $p_t$, $\eta$, and centrality. $v_6$,
although essentially zero, is not inconsistent with scaling as
$v_2^3$. This is the first measurement of higher harmonics at RHIC and
it is expected that these higher harmonics will be a sensitive test of
the initial configuration of the system, since they provide a Fourier
analysis of the shape in momentum space which can be related back to
the initial shape in configuration space. In fact, it has been
emphasized that $v_4$ has a stronger potential than $v_2$ to constrain
model calculations and carries valuable information on the dynamical
evolution of the system.

%===========================================================================
\begin{acknowledgments}
{\it Acknowledgments---} We wish to thank Jean-Yves Ollitrault and
Peter Kolb for extensive discussions. We thank the RHIC Operations
Group and RCF at BNL, and the NERSC Center at LBNL for their
support. This work was supported in part by the HENP Divisions of the
Office of Science of the U.S.  DOE; the U.S. NSF; the BMBF of Germany;
IN2P3, RA, RPL, and EMN of France; EPSRC of the United Kingdom; FAPESP
of Brazil; the Russian Ministry of Science and Technology; the
Ministry of Education and the NNSFC of China; SFOM of the Czech
Republic, DAE, DST, and CSIR of the Government of India; the Swiss
NSF.
\end{acknowledgments}

%===============================================================================

\def\etal{\mbox{$\mathrm{\it et\ al.}$}}

%============================================================================
\end{document}